\documentclass[a4paper]{jpconf} 
\usepackage[a4paper,width=150mm,top=40mm,bottom=27mm,bindingoffset=6mm]{geometry}
\usepackage{amsmath}
\usepackage{amssymb}
\usepackage{amsthm}
 
\newtheorem{theorem}{Theorem}[section]
\newtheorem*{theorem*}{Theorem}

\newtheorem{proposition}[theorem]{Proposition}
\theoremstyle{definition}
\newtheorem{definition}[theorem]{Definition}
\theoremstyle{corollary}

\theoremstyle{remark}

\theoremstyle{conclusion}

\usepackage{abstract}

\begin{document}
\title{On polynomial symmetry algebras underlying superintegrable systems in Darboux spaces}

\author{ Ian Marquette \footnote{i.marquette@uq.edu.au},  Junze Zhang\footnote{junze.zhang@uqconnect.edu.au}  and Yao-Zhong Zhang\footnote{yzz@maths.uq.edu.au}}

\address{School of Mathematics and Physics, The University of Queensland, Brisbane, QLD 4072, Australia}


\begin{abstract}
\noindent  
We review three different approaches to polynomial symmetry algebras underlying superintegrable systems in Darboux spaces. The first method consists of using deformed oscillator algebra to obtain finite-dimensional representations of quadratic algebras. This allow one to gain information on the spectrum of the superintegrable systems. The second method has similarities with the induced module construction approach in the context of Lie algebras and can be used to construct infinite dimensional representations of the symmetry algebras. Explicit construction of these representations is a non-trivial task due to the non-linearity of the polynomial algebras. This method allows the construction of states of the superintegrable systems beyond the reach of separation of variables. As a result, we are able to construct a large number of states in terms of Airy, Bessel and Whittaker functions which would be difficult to obtain in other ways. We also discuss the third approach which is based on the notion of commutants of subalgebras in the enveloping algebra of a Poisson algebra or a Lie algebra. This allows us to discover new superintegrable models in the Darboux spaces and to construct their integrals and symmetry algebras via polynomials in the enveloping algebras. 
\end{abstract}

\section{Introduction}

Finite and Infinite-dimensional representations of symmetry algebras play a significant role in determining the spectral properties of physical Hamiltonians.
In \cite{Marquette:2023wxn,marquette2023infinite} we focused on the representations of polynomial symmetry algebras underlying superintegrable systems in 2D Darboux spaces. As a result, we are able to construct a large number of states in terms of the Airy, Bessel and Whittaker functions which would be difficult to obtain in other ways.

Let $(\mathcal{M},g_{ij})$ be a smooth manifold with a metric tensor over $\mathbb{C}$. Suppose that $\mathcal{M}$ admits local separable coordinates  $(x_1,\ldots,x_n)$.  Let \begin{align}
      \hat{\mathcal{H}} = \sum_{i,j=1}^n \frac{1}{\sqrt{\det (g_{ij})}} \dfrac{\partial }{\partial x_j} \left( \sqrt{\det (g_{ij})} g_{ij} \dfrac{\partial  }{ \partial x_j} \right) + V(x_1,\ldots,x_n)  \in \Gamma\left(T^*\mathcal{M}\right) \label{eq:Ham}
\end{align}
be the Hamiltonian of a superintegrable system in $(\mathcal{M},g_{ij})$  and $S_m = \{  \hat{\mathcal{H}},\hat{X}_1,\ldots,\hat{X}_m\}$ be a set of integrals of motion.   Let $\hat{\mathcal{Q}}(d)$ denote the polynomial algebra of order $d$ over the polynomial ring $\mathbb{C}[\hat{\mathcal{H}}]$,  generated  by the integrals from $S_m$ with the following non-trivial commutators \begin{align}
    [\hat{X}_s,\hat{X}_t] = \sum_q f_q( \hat{\mathcal{H}}) \hat{X}_q + \sum_{p,q} f_{p,q}(\hat{\mathcal{H}}) \hat{X}_p \hat{X}_q +\sum_{p,q,r} f_{p,q,r}(  \hat{\mathcal{H}}) \hat{X}_p \hat{X}_q \hat{X}_r + \ldots.
    \label{eq:aba}
\end{align} Here $f_{p,q,r}(  \hat{\mathcal{H}})$ is a polynomial function of the Hamiltonian $\hat{\mathcal{H}}.$ Let $V$ be a vector space and  $\mathfrak{gl}(V)$ be the space of endomorphism of $V.$ Then representations of the associative polynomial algebra $\hat{\mathcal{Q}}(d)$ are given by $\rho: \hat{\mathcal{Q}}(d) \rightarrow \mathfrak{gl}(V)$.


 
\section{Construction of representations of polynomial algebras}
\label{b}

Superintegrable systems in 2D Darboux spaces were classified \cite{MR2023556,MR1878980} and it was found that there exist 12 distinct classes of second order superintegrable systems in the Darboux spaces. In \cite{Marquette:2023wxn} we presented exact solutions via purely algebraic means for the energies of all the 12 classes of superintegrable systems in four different 2D Darboux spaces. This was achieved by constructing the deformed oscillator realization and finite-dimensional irreducible representation of the underlying quadratic symmetry algebra generated by quadratic integrals respectively for each of the 12 superintegrable systems.  
 
\subsection{Deformed oscillator algebra realizations}
\label{2.2}
One way to construct representations of quadratic symmetry algebras is through their realizations in terms of the deformed oscillator algebra. 

Consider the generic quadratic algebra $\hat{\mathcal{Q}}(3)$ with three generators, proposed in \cite{MR1814439,MR2226333}, \begin{align}
\begin{matrix}
  [A,B] = C,\\
    [A,C] = \alpha A^2 + \gamma \{A,B\} + \delta A + \epsilon B + \zeta ,\\
     [B,C] = a A^2 - \gamma B^2 -\alpha \{A,B\} + dA - \delta B + z.   
\end{matrix} \label{eq:comm}
\end{align} Its Casimir operator is given by \begin{align*}
    K  = & C^2 - \alpha \{A^2,B\} - \gamma \{A,B^2\} + (\alpha \gamma -\delta)\{A,B\} + (\gamma^2 -\epsilon)B^2 + (\gamma \delta - 2 \zeta) B  \\
    & + \frac{2a}{3} A^3 + (d + \frac{a \gamma }{3} + \alpha^2) A^2 + (\frac{a \epsilon}{3} + \alpha \delta + 2z) A.
\end{align*} 
Notice that the generic form $\eqref{eq:comm}$ was proposed for the case of two quadratic integrals $A, B$ and is quite formal because in general there is no guarantee that integrals of degree 2 would close to form a quadratic algebra.   

Finite-dimensional representations of $\eqref{eq:comm}$ can be conveniently constructed via the deformed oscillator algebra with generators $\{b^\dagger,b,\mathcal{N}\}$, which denotes by $\mathfrak{o}(3),$ satisfying the following commutation relations \cite{MR1814439} \begin{align}
    [\mathcal{N},b^\dagger] = b^\dagger, \quad\text{ } [\mathcal{N},b] = -b, \quad\text{ } b b^\dagger  = \Phi(\mathcal{N} + 1),\quad\text{ } b^\dagger b = \Phi(\mathcal{N}). \label{eq:alg}
\end{align} Here $\mathcal{N}$ is the number operator and $\Phi$ is a well defined real function satisfying 
\begin{align}
    \Phi(0) = 0, \text{ } \Phi(x) > 0, \text{ for all } x \in \mathbb{R}^+. \label{eq:consta}
\end{align}  Then representations of $\mathfrak{o}(3)$ are given by
\begin{align}
    \mathcal{N} \psi_n = n \psi_n,\quad \text{ } b^\dagger \psi_n = \sqrt{\Phi(n+1)} \psi_{n+1}, \quad  b \psi_n = \sqrt{\Phi(n)} \psi_{n-1}. \label{eq:}
\end{align} where $\psi_{n}$ are eigenvectors that form the Fock basis of the oscillator algebra. 
Imposing $\Phi(p+1) = 0 $ for all $p \in \mathbb{N}$, one obtains $p+1$-dimensional unitary representations of 
$\mathfrak{o}(3)$. 

Applying the deformed oscillator technique, in \cite{Marquette:2023wxn} we gave algebraic derivations of the spectra for the 12 superintegrable systems in the 2D Darboux spaces. As an example, we here review some of the results for the superintegrable system in the Darboux space II with the Hamiltonian $\hat{\mathcal{H}}  = \frac{x^2}{x^2 + 1}   \left( \partial_x^2 + \partial_y^2\right) +  \frac{x^2}{x^2 + 1} \left( a_1 \left( \frac{x^2}{4} + y^2 \right) + a_2 y + \frac{a_3}{x^2} \right)$.
The constants of motion are  $  A = \partial_y^2 + a_1y^2 + a_2y $ and  \begin{align*}
    B  = \frac{2 y}{x^2 + 1}  \left( \partial_y^2 - x^2 \partial_x^2 \right)  + 2x \partial_x \partial_y + \partial_y + \frac{a_1}{2} y \left(x^2 + \frac{x^2 + 4 y^2}{x^2 + 1}\right) + \frac{a_2}{2} \left(x^2 + \frac{ 4 y^2}{x^2 + 1}\right) - \frac{2 a_3 y}{x^2 +1}.
\end{align*} 
These integrals satisfy the quadratic algebra $\mathcal{Q}(3)$   \cite{MR2023556}  \begin{align*}
    & [A,B] = C, \quad  [A,C] =  - 4 a_1 B - 4a_2 A,\\
    & [B,C] = - 24 A^2 + 4 a_2 B + 32 \hat{\mathcal{H}} A - 8 \hat{\mathcal{H}}^2 - 8a_1 \hat{\mathcal{H}} + 6 a_1 + 8 a_1 a_3.
\end{align*}   with the Casimir operator given by \begin{align*}
    K  = C^2 - 16 A^3 + 4 a_1 B^2 + 4 a_2\{A,B\}+ \left(4 a_1( 4 a_3 - 11)  - (16 a_1 \hat{\mathcal{H}}+ 16  \hat{\mathcal{H}}^2 ) \right) A + 32 \hat{\mathcal{H}}A^2.   
\end{align*} 
In term of the differential realization of $A, B$,  the Casimir $K$ takes the simple form   $K= (32 a_1 +4 a_2^2) \hat{\mathcal{H}} - a_2^2  (3 +4 a_3 )$. It was shown in \cite{Marquette:2023wxn} that the transformation $A = 2 \sqrt{-a_1} (\mathcal{N} + \eta), \quad \text{ } B = \frac{2a_2}{\sqrt{-a_1}}(\mathcal{N} + \eta) + \frac{a_2 \hat{\mathcal{H}}_{2,1}}{a_1} + b^\dagger + b,  $  maps $\hat{\mathcal{Q}}(3)$ is to the deformed oscillator algebra $\eqref{eq:alg}$ with the structure function cubic in $(\mathcal{N} + \eta)$



Here $\eta$ is a constant determined from the constraints of the structure function. 
In general it is very difficult to obtain analytical solutions to $\eta$ for arbitrary coefficients $a_i$. We consider the case where $-a_1=a_2=a_3=a$, $a\in\mathbb{R}$. For such model parameters, the energies $E$ and their corresponding structure functions for distinct $\eta$  are \begin{align}
    E_\epsilon  = \frac{1}{4} \left(8 \sqrt{a} (p+1)+3 a+  2 \epsilon\sqrt{8 a^{3/2} (p+1)-2 a^2+a} \right),
\end{align} 
where $\epsilon = \pm 1$, with the associated structure function $\Phi^{(II)}_{E_\epsilon}(z) = z(p+1 - z)^2.$ The energy spectrum $E_\epsilon$ is real for $0 < a \leq 1/2$.   The corresponding energy spectrum of the system and structure function for the $(p+1)$-dimensional unirreps of the deformed oscillator algebra are given respectively by $ E = p(p+2)+a+\frac{3}{4} $ and
\begin{align}   
\Phi_E (z) = z(z-p-1)\left(z+  \frac{1}{8 a}\left(3 a^{3/2}- 4 a (p+1) +\sqrt{a} (4 p^2+8 p+3)\right) \right). 
\end{align} 

\subsection{Verma module constructions on $\hat{\mathcal{Q}}(d)$}
 
In this subsection, we review our method in \cite{marquette2023infinite} of determining the representations of $\hat{\mathcal{Q}}(d)$ without relying on the deformed oscillator algebra realizations. In what follows, we always assume that the Schr{\"o}dinger equation $\hat{\mathcal{H}} \Psi= E \Psi$ has solutions of the separable form $\Psi(x_1,\ldots,x_n) =F_1(x_1) \ldots F_n(x_n)$, where $\hat{\mathcal{H}}$ is the Hamiltonian defined in $\eqref{eq:Ham}$ and $F_j$, $j=1,2,\ldots,$ are functions of the coordinates $x_j$. Notice that the solution space of the Schr{\"o}dinger equation will form an infinite-dimensional vector space. As $\hat{\mathcal{Q}}(d)$ is the symmetry algebra of the Hamiltonian $\hat{\mathcal{H}}$,  infinite dimensional representations of $\hat{\mathcal{Q}}(d)$ can be obtained through actions of its generators on eigenstates of $\hat{\mathcal{H}}$.  
 
Since solutions of the Schr{\"o}dinger equation are separable, we assume that $S_p = \{\hat{\mathcal{H}},\hat{X}_1,\ldots,\hat{X}_p\}$ is a subset of $S_m$ such that $\hat{X}_j \Psi = \lambda_j \Psi$, with $\lambda_j \in \mathbb{R}$ for all $ 1 \leq j \leq p.$ It is clear that the existence of $S_p$ depends on the explicit form of the integrals. Moreover, superintegrable systems with scalar potentials and $n$ degrees of freedom admit up to $n$ integrals (including the Hamiltonian) forming a set of commuting operators (then $p \leq n$).  
Then for some $n_j \in \mathbb{N}$ define the following reiterated vector\begin{align}
   \prod_{p+1 \leq j \leq m} \hat{X}_{j}^{n_j} \Psi = \Psi_{n_{p+1} ,\ldots,n_m }, \label{eq:a1}
\end{align}  
in the eigenspace $V_E$ of $\hat{\mathcal{H}}$, i.e. 
$  \Psi_{n_{p+1} ,\ldots,n_m } \in V_E = \{\textbf{v} : \hat{\mathcal{H}} \textbf{v} = E \textbf{v}\}.$   Due to the complexities of the commutation relations $\eqref{eq:aba}$ and the form of $\Psi$, analytic computation of $\prod_{p+1 \leq j \leq m} \hat{X}_{j}^{n_j}$ on $\Psi$ is not in general feasible. 
 
From \cite[\text{Theorem 1}]{MR0460751}, actions by any elements in $S_p$ on $\Psi_{n_{p+1} ,\ldots,n_m }$ are still in the eigenspace space.  For integrals $\hat{X}_i \in S_m/S_{p},$ we aim to construct the recurrence relations 
\begin{align}
 \hat{X}_i \Psi_{n_{p+1} ,\ldots,n_m } = \sum_{i_1 \in W_1,\,\ldots,\,i_m \in W_m}   \Psi_{i_1,...,i_m}, \label{eq:recur}
\end{align}
where $W_j$ are the sets of integer tuples and $S_m/S_p$ defines a subset of $S_m$ with all the elements in $S_p$ excluded. This allows us to understand how the operators in $S_m/S_p$ acts on the eigenspace of $S_p$. Let $$V_{n_j}: = \{\Psi, X_{p+1} \Psi,\ldots,\Psi_{n_{p+1}},\ldots,X_{p+2} \Psi_{n_{p+2}},\ldots,  \Psi_{n_{p+1},\ldots,n_m}\}.$$ In general it is not easy to show that the elements in $V_{n_j}$ form a basis. However, if $V_{n_j}$ is a finite-dimensional vector space of homogeneous polynomials, then the infinite-dimensional representations of $\hat{\mathcal{Q}}(d)$ are given by $\pi: \hat{\mathcal{Q}}(d) \rightarrow \mathfrak{gl}(V)$ with $V= \bigoplus_{n_j \in \mathbb{N}} V_{n_j}.$ The polynomial commutation relations $\eqref{eq:aba}$ can be used to simplify the vectors in $\eqref{eq:a1}$ for explicit construction of the recurrence relations $\eqref{eq:recur}$. 

As illustrations, we apply the above construction to the cubic algebra underlying the superintegrable integrable system $S_3 = \{\hat{\mathcal{H}}, \hat{X}_1,\hat{X}_2\}$ in the 2D Darboux space with separable local coordinates $(x,y)$. Here $\hat{\mathcal{H}} = \varphi(x) \left(\partial_x^2 + \partial_y^2 + c\right)$,  where $c$ is a constant, is the Hamiltonian and $\hat{X}_1, \hat{X}_2$ are integrals. Solution of the Schr{\"o}dinger equation has the separable form  $\Psi(x,y) = X(x)Y(y)$, where $X(x)$ and $Y(y)$ satisfy the second order ODE $  X'' + \lambda \varphi(x) X = 0$ and $ Y''- \lambda Y = 0$ with separation constant $\lambda.$  As shown in \cite{Marquette:2023wxn}, the integrals form cubic algebra $\hat{\mathcal{Q}}(3)$ with the following commutation relations  
\begin{align}
  [\hat{X}_1,\hat{X}_2] =&\hat{F}, \nonumber \\
  [\hat{X}_1,\hat{F}]  =& u_1 \hat{X}_1^2 + u_2 \hat{X}_1 + u_3\hat{X}_2 +  u, \label{eq:a}\\
  [\hat{X}_2,\hat{F}]= & v_1 \hat{X}_1^3 +v_2 \hat{X}_1^2 + v_3 \hat{X}_1-u_2 \hat{X}_2 -  u_1\{\hat{X}_1,\hat{X}_2\} + v,\nonumber   
\end{align}  where $u_j,v_j,\ldots,u,v$ are polynomials of the Hamiltonian $\hat{\mathcal{H}}$ and  $v_1$ in (\ref{eq:a}) is non-zero coefficient. The Casimir operator $C_{(3)}$ for the cubic algebra is given by \begin{align*}
    C_{(3)}  = & \hat{F}^2 - u_1 \{\hat{X}_1^2,\hat{X}_2\}   -u_2\{\hat{X}_1,\hat{X}_2\} + \frac{v_1}{2} \hat{X}_1^4+\frac{2}{3}v_2 \hat{X}_1^3 \\
    & + \left(  v_3 + u_1^2  \right) \hat{X}_1^2 + \left(u_1u_2   + 2v \right) \hat{X}_1 
    - 2u \hat{X}_2  - u_3 \hat{X}_2^2 .
\end{align*}  For the current case, the subset $S_1 = \{\hat{\mathcal{H}},\hat{X}_1\} \subset S_3$ contains the simultaneously diagonalizable operators. We can show that the set $V_{m,n}$ formed by the eigenvectors $\hat{X}_2^n\hat{F}^m$ is a vector space such that $\rho: \hat{\mathcal{Q}}(3) \rightarrow \mathfrak{gl}(V  )$ provides infinite dimensional representations, where $V  =\oplus_{m,n\in \mathbb{N}} V_{m,n} $ and $V_{m,n} = \{\Psi,\ldots, \hat{F}^m\Psi, \ldots, \hat{X}_2^n \hat{F}^m \Psi\}.$
Extending results proved in \cite[Proposition 3.1, Proposition 3.4]{marquette2023infinite}, we state
\begin{proposition}
    Let $\hat{\mathcal{Q}}(3)$ be the cubic algebra $\eqref{eq:a},$ and let $W$ be the space cyclically generated by $\hat{F}.$ Let $\psi_m = \hat{F} \Psi^m$ for all $m \in \mathbb{N}$. Suppose that $ u_3 =0$. Then $V \cong W$ and $V$ form a $\hat{\mathcal{Q}}(3)$-module. In particular, if $u_1 = u_2 = u_3= 0,$ then the representation of $\rho: \hat{\mathcal{Q}}(3) \rightarrow \mathfrak{gl}(W)$ has the form of 
    \begin{align*}
        \hat{F}\psi_m = & E \psi_{m+1},\\
        \hat{X}_1 \psi_m = & (m-2) u \psi_m + \lambda \psi_m ,\\
        \hat{X}_2 \psi_m = & f(\psi_{m-3},\psi_{m-2},\psi_{m-1},\psi_{m},\psi_{m+2})
    \end{align*} 
where $f$ has coefficient depending on certain polynomials of $m$.

\end{proposition}
For further details, we refer the reader to the paper.

\section{Polynomial algebras with Poisson-Lie brackets under PBW basis}
\label{c}

Let's first define the commutants relative to subalgebras of the universal enveloping algebra and the symmetry algebra of a Lie algebra $\mathfrak{g}.$

\begin{definition}
\label{2.1}
Let $\mathfrak{g}$ be a $n$-dimensional Lie algebra with a basis $\beta_\mathfrak{g} = \{X_1,\ldots,X_n\}$, and let $\mathfrak{g}^*$ be its dual with a basis $\beta_{\mathfrak{g}^*} = \{x_1,\ldots,x_n\}$.  Let $(\mathcal{U}(\mathfrak{g}),[\cdot,\cdot]$ and $(\mathcal{S}(\mathfrak{g}^*),\{\cdot,\cdot\})$ be the universal enveloping algebra of $\mathfrak{g}$ and the symmetric algebra of $\mathfrak{g^*}$, respectively. Then the $\textit{commutants}$ relative to subalgebras $ \mathfrak{a} \subset \mathfrak{g}$ and $\mathfrak{a}^* \subset \mathfrak{g}^*$ in $\mathcal{U}(\mathfrak{g})$ and $\mathcal{S}(\mathfrak{g}^*)$, denoted respectively as $\mathcal{C}_{\mathcal{U}(\mathfrak{g})}(\mathfrak{a})$   and $\mathcal{C}_{\mathcal{S}(\mathfrak{g}^*)}(\mathfrak{a}^*)$, are defined as follows
\begin{align*}
      \mathcal{C}_{\mathcal{U}(\mathfrak{g})}(\mathfrak{a}) &= \left\{ Y \in \mathcal{U}(\mathfrak{g}): [X,Y] = 0, \quad \forall X \in \mathfrak{a}\right\},\\
      \mathcal{C}_{\mathcal{S}(\mathfrak{g}^*)}(\mathfrak{a}^*)  & = \{y \in \mathcal{S}(\mathfrak{g}^*): \{x,y\} = 0, \quad \forall x \in \mathfrak{a}^*\}.
  \end{align*}
\end{definition} 

The adjoint action of  $\mathfrak{g}$ and co-adjoint action of  $\mathfrak{g}^*$ on the universal enveloping algebra $\mathcal{U}(\mathfrak{g})$ and the symmetric algebra $\mathcal{S}(\mathfrak{g}^*)$ are given by \begin{align}
      P\left( X_1,\ldots,X_n\right) \in \mathcal{U}(\mathfrak{g}) &\mapsto  [X_j,P] \in \mathcal{U}(\mathfrak{g}) , \\
      p(x_1,\ldots,x_n) \in \mathcal{S}(\mathfrak{g}^*) & \mapsto \{x_j,p  \} = \tilde{X}_j(p(\textbf{x})) = \sum_l C_{jk}^lx_l \dfrac{\partial p}{\partial x_j} \in \mathcal{S}(\mathfrak{g}^*), \label{eq:dual}
\end{align}  
respectively, where $\tilde{X}_j = \sum_l C_{jk}^l x_l \dfrac{\partial}{\partial x_j}$ are vector field realizations of the generators of the Lie algebra $\mathfrak{g} $ and $\{\cdot,\cdot\} : C^\infty(\mathfrak{g}^*) \times C^\infty(\mathfrak{g}^*) \rightarrow C^\infty(\mathfrak{g}^*)$ is a Poisson-Lie structure induced by the co-adjoint action of $\mathfrak{g}^*$. 
  
Let us provide a short review on the construction  of symmetry algebras from the subalgebras $\mathfrak{a}$ of $\mathfrak{g}$ (see also \cite{MR1520346,MR191995}). Using $\eqref{eq:dual},$  the commutant $\mathcal{C}_{\mathcal{S}(\mathfrak{g}^*)}(\mathfrak{a}^*)$ is generated by the linearly independent solutions of the  systems of PDEs \begin{align}
    \tilde{X}_j(p_h)(\textbf{x}) = \{x_j,p_h\}(\textbf{x}) = \sum_l C_{jk}^lx_l \dfrac{\partial p_h}{\partial x_j} = 0, \text{ }\quad 1 \leq j \leq \dim  \mathfrak{a} \equiv s, \label{eq:func}
\end{align}
where $p_h(\textbf{x})$ is a homogeneous polynomial of degree $h \geq 1$ with the generic form\begin{align}
    p_h(\textbf{x}) = \sum_{i_1 + \ldots + i_s \leq h} \Gamma_{i_1,\ldots,i_s}\, x_1^{i_1} \ldots x_s^{i_s} \in \mathcal{S}(\mathfrak{g}^*). \label{eq:ci}
\end{align} 
Notice that, depending on the structure of the Lie algebra $\mathfrak{g},$ solutions of $\eqref{eq:func}$ may not be polynomials ( see \cite{campoamor2023polynomial} and the reference therein). 

By systematically analyzing the polynomial solutions of $\eqref{eq:func} $ up to certain degrees, we obtain polynomials that can be decomposed in terms of products of polynomials of lower degrees. 
Using the symmetrization map, commutants in the enveloping algebra are obtained as $\mathcal{C}_{\mathcal{U}(\mathfrak{g})}(\mathfrak{a}) = \Lambda \left(\mathcal{C}_{\mathcal{S}(\mathfrak{g}^*)}(\mathfrak{a}^*)\right)$. 
Then the linearly independent monomials form a finite-generated Poisson algebra.
From the construction above, one can define algebraic Hamiltonians as follows: 


 
\begin{definition}
\label{H} \cite{MR2515551}.
Let $\mathfrak{a} \subset \mathfrak{g}$ be a Lie subalgebra and $\mathcal{C}_{\mathcal{U}(\mathfrak{g})}(\mathfrak{a})$ be its commutant, where $\mathfrak{a}$ admits a basis $\beta_\mathfrak{a} = \{X_1,\ldots,X_s\} $ with $\dim\mathfrak{a} =s.$ An algebraic Hamiltonian  with respect to $\mathfrak{a}$ is given by \begin{align*}
      \hat{\mathcal{H}} = \sum_{1 \leq i_1,\ldots,i_k \leq h}^s \Gamma_{i_1,\ldots,i_k}X_{i_1}\ldots X_{i_k}  +  \sum_t C_t K_t,
\end{align*} where $\Gamma_{i_1,\ldots,i_k} $ is the constant coefficient and $K_t$ are the Casimir invariants of $\mathfrak{g}$.
\end{definition} 



In the following we obtain polynomial algebras from the 2D $\textit{conformal algebra}$ $\mathfrak{c}(2) $ and their representations.  In  \cite{MR3988021}, the connection of $\mathfrak{c}(2)$ with 2D Darboux spaces was established. As an example we here present a derivation of an algebraic Hamiltonian and the underlying symmetry algebra from the conformal algebra.
%


We consider the subalgebra $\mathfrak{a}_1$ and construct the commutant $\mathcal{C}_{\mathcal{S}(\mathfrak{c}^*(2))}(\mathfrak{a}_1^*)$. By solving $   \left\{p(x_1,\ldots,x_6),x_1\right\}  = 0, $ we find $6$ linearly independent polynomials as follows  $   A_1 =  x_1,  $ $ A_2 = x_2, $ $ A_3 =x_1 x_3 + x_2 x_4,$ $A_4 = x_2 x_6 + x_3^2 ,$ $ A_5 = x_1x_6-2x_3 x_4-x_2x_5,$ $ A_6 = - x_1 x_5 +  x_4^2.$ Then possible algebraic Hamiltonian is given by $$ \mathcal{H}  = \alpha x_1 + \gamma_1 \mathcal{C}_1 + \gamma_2 \mathcal{C}_2. $$ 
It is easy to check that $\{\mathcal{H},A_j\} = 0$ for $1 \leq j \leq 6.$ The linearly independent polynomial integrals form a quadratic Poisson algebra $\mathcal{Q}_1(2) =\mathcal{Q}_1 \oplus \mathcal{Q}_2$, where $\mathcal{Q}_1 =\mathrm{Span}\{A_1,A_2\}$ and $\mathcal{Q}_2 = \mathrm{Span}\{A_3,\ldots,A_6\},$ with the following non-zero brackets \begin{align}
   \{A_2,A_3\} & = A_1^2 + A_2^2,\quad \text{ } \{A_2,A_4\} =  -\{A_2,A_6\} = 2 A_3 , 
   \nonumber\\
     \{A_3,A_4\} &=- A_1 A_5 -2 A_2 A_6 = -  \{A_3,A_6\}  .   \label{eq:} 
\end{align}   Notice that $\mathcal{Z}\left(\mathcal{Q}_1(2)\right) =\mathrm{Span}\{A_1,A_5\}.$  The universal enveloping Poisson algebra has the form \cite{MR3659329} 
$$K_{\mathcal{Q}_1(2)}^h =\sum_{\alpha_1+ \ldots + \alpha_6 \leq h} \Gamma_{\alpha_1,\ldots,\alpha_6} A_1^{\alpha_1} \ldots A_6^{\alpha_6} \in \mathcal{S}_h\left(\mathcal{Q}_1(2)\right),$$ where $\mathcal{S}_h(\cdot)$ is the symmetry algebra of the finitely-generated quadratic Poisson algebra $\mathcal{Q}_1(2).$ 
The functionally independent Casimir operators are \begin{align*}
   & K_{\mathcal{Q}_1(2)}^{1,1} = A_1, \quad\text{ } K_{\mathcal{Q}_1(2)}^{1,2}= A_5, \quad\text{  }   K_{\mathcal{Q}_1(2)}^{1,3} = A_4 + A_6,\\
   & K_{\mathcal{Q}_1(2)}^{3,1} = (A_1^2+A_2^2)A_6+A_2 A_5 A_1+A_3^2.
\end{align*}
  
The corresponding commutator algebra generated by the integrals has the following
commutation relations \begin{align*}
   [\hat{A}_2,\hat{A}_3] & = \hat{A}_1^2 + \hat{A}_2^2,\quad \text{ } [\hat{A}_2,\hat{A}_4] = -[\hat{A}_2,\hat{A}_6]  = 2 \hat{A}_3 , \\
   [\hat{A}_3,\hat{A}_4] & = 2\left( \hat{A}_2 \hat{A}_6 - \hat{A}_3 \right) + 2 \hat{A}_1 \hat{A}_5 =  
   [\hat{A}_3,\hat{A}_6]   . 
 \end{align*} 
It is clear that this is a quadratic algebra.  

\section{conclusion}

 In this short note, we have reviewed three different approaches to the construction of polynomial symmetry algebras of superintegrable systems and their representations. We have first described the application of the deformed oscillator algebra technique to the superintegrable system in the Darboux space. Then we have described the method for constructing infinite-dimensional representations and commented its relevance to solving superintegrable systems without the use separation of variables. Finally, we have described the algebraic method based on commutants and as an example have applied the construction to obtain the algebraic Hamiltonian and underlying quadratic symmetric algebra from the 2D conformal algebra.

\section*{Acknowledgement}

IM was supported by the Australian Research Council Future Fellowship FT180100099. YZZ was supported by the Australian Research Council Discovery Project DP190101529. 

\section{Bibliography}
  
\bibliographystyle{unsrt}   
\bibliography{bibliography.bib}
\end{document}